\documentclass[12pt]{article}
\usepackage[dvips]{graphicx}

\newcommand{\sph}	{{\rm sph}}

\newcommand{\rad}	{{\rm rad}}
\newcommand{\external}	{{\rm ext}}
\newcommand{\internal}	{{\rm int}}

\newcommand{\diag}	{{\rm diag}}
\newcommand{\BG}	{{\rm BG}}
\newcommand{\vac}	{{\rm vac}}

\newcommand{\BH}	{{\rm BH}}
\newcommand{\pl}	{{\rm pl}}
\newcommand{\maxi}	{{\rm max}}
\newcommand{\mini}	{{\rm min}}

\newcommand{\rhopeak}	{{\rho\mbox{-}{\rm peak}}}

\newcommand{\weak}	{{\rm W}}
\newcommand{\GUT}	{{\rm GUT}}

\newcommand{\lnear}	{{\begin{array}{c} < \\[-1.1em] \sim \end{array}}}
\newcommand{\gnear}	{{\begin{array}{c} > \\[-1.1em] \sim \end{array}}}
\newcommand{\GeV}	{{\rm \;GeV}}

\newcommand{\EQ}[1]	{(\ref{#1})}

\begin{document}

\begin{flushright}
 \begin{minipage}[b]{43mm}
  hep-th/0312060\\
  WIS/33/03-DEC-DPP\\
 \end{minipage}
\end{flushright}

\renewcommand{\thefootnote}{\fnsymbol{footnote}}
\begin{center}
 {\Large\bf
 Black Hole as a Baryon-Reactor ---\\
 Rapid Baryon Number Violation in Black Hole
 }\\
 \vspace*{3em}
 {Yukinori Nagatani}\footnote
 {e-mail: yukinori.nagatani@weizmann.ac.il}\\[1.5em]
 {\it Department of Particle Physics,\\
 The Weizmann Institute of Science, Rehovot 76100, Israel}
\end{center}
\vspace*{1em}

\begin{abstract}
 We find out that baryon numbers of the matter fallen into black holes
 are rapidly washed-out
 by investigating the radiation-ball description of the black holes.
 The radiation-ball solution, which was derived by analyzing
 the backreaction of the Hawking radiation into space-time
 and is identified as a black hole,
 consists of the radiation gravitationally-trapped into the ball
 and of a singularity.
 The baryon number of the black hole is defined as
 that of the radiation in the ball.
 The sphaleron processes of the Standard Model work in the ball
 because the proper temperature of the radiation is Planck scale
 and the Higgs vev becomes zero.
 The decay-rate of the baryon number becomes
 $\dot{B}/B \approx -\alpha_\weak^4 / r_\BH$
 for the Schwarzschild black hole of radius $r_\BH$.
 When we assume the baryon number violating processes of the GUT,
 we find more rapid decay-rate
 $\dot{B}/B \approx -\alpha_\GUT^2 / r_\BH$.
 We can regard the black holes as the baryon-reactors
 which convert the baryonic matter into energy of radiation.
\end{abstract}

%

\newpage
\section{Introduction}\label{intro.sec}

Whether information of the matters which have fallen into black holes
is conserved or not is one of the most important problems
on the black hole physics.
Especially baryon numbers of the black holes are quite interesting
because the solar-massive black holes created by the supernova
as a consequence of star-evolutions consist from the baryonic matter
and seem to have huge baryon numbers.
The Hawking radiations do not radiate the net baryon number
because they are thermal in the ordinary arguments
\cite{Hawking:1975sw,Hawking:1974rv}.
Are the baryon numbers lost, do the Hawking radiations carry out or
are remnants carrying the baryon number
left after the completion of the radiation?

In this paper we consider this problem by investigating the
radiation-ball description of the black holes.
The radiation-ball was derived by solving the Einstein equation
including the backreaction of the Hawking radiation into space-time
by using the proper temperature ansatz 
\cite{Nagatani:2003rj,Nagatani:2003new}.
The ball consists of radiation trapped into the ball by deep
gravitational potential and of a repulsive singularity.
The balls are identified as black holes with quantum mechanical
properties including the Hawking radiation
\cite{Hawking:1975sw,Hawking:1974rv}
and the Bekenstein entropy
\cite{Bekenstein:1973ur}.
The radius of the ball equals to
the horizon radius of the corresponding black hole.
We understand that the horizon vanishes by the backreaction and
the radiation-ball appears.
The Hawking radiation is regarded as a leak-out of the radiation
from the ball.
The Bekenstein entropy is considered as being carried
by the inside radiation
because the total entropy of the ball reproduces the area-law
and is near the Bekenstein entropy.
There arises no information paradox because there is no horizon.
The inside radiation of the ball has Planck temperature.

The picture of the radiation-ball allows us to analyze
the time-evolution of the baryon number of the black hole concretely
because the baryon number of the black hole is simply carried by
the inside radiation of the ball and
the radiation has finite temperature.
The Higgs vacuum expectation value (vev) in the ball is zero
and the electroweak (EW) symmetry is restored
because the inside temperature is much higher than the EW scale.
The baryon number in the ball is not conserved
because the sphaleron processes work in the ball.
The time-evolution of the baryon number is computed
by the Boltzmann equation.

We find that the baryon number of the black hole exponentially decreases
with the decay-rate $\dot{B}/B \approx -\alpha_\weak^4 / r_\BH$,
where $r_\BH$ is radius of the Schwarzschild black hole and
$\alpha_\weak$ is the weak gauge coupling constant.
The time scale of the decay-rate is much shorter than the lifetime of
the black hole.
The time scale becomes order of several seconds
for the solar-massive black hole.
We have omitted effects of the grand unified theory (GUT).
If we take account the baryon-number violating interactions in the GUT,
we expect more rapid rate $\dot{B}/B \approx -\alpha_\GUT^2 / r_\BH$.
Therefore the baryon number of the black holes is rapidly washed-out
and we can regard the black holes as the baryon-reactors
which convert the baryonic matter into energy of radiation.

The paper is organized as follows.
In the next section 
we review the radiation-ball solution.
In Section \ref{baryon.sec},
we consider the time-evolution of the baryon number.
In the final section we provide a conclusion and discussions.

\section{Radiation Ball}\label{radiationball.sec}

We briefly review the radiation-ball solution,
which is a spherically symmetric static solution of
the Einstein equation with the radiation
obeying the local proper temperature ansatz
\cite{Nagatani:2003rj,Nagatani:2003new}.
The generic metric of  the spherically symmetric static space-time
parameterized by the time coordinate $t$
and the polar coordinates $r$, $\theta$ and $\varphi$ is
\begin{eqnarray}
 ds^2
  &=&
  F(r) dt^2 - G(r) dr^2 - r^2 d\theta^2 - r^2 \sin^2\theta \, d\varphi^2.
  \label{generic-metric}
\end{eqnarray}
The unknown functions $F(r)$ and $G(r)$ are determined by the Einstein
equation.
We will consider the situation of the equilibrium
between a Schwarzschild black hole with Hawking radiation
and a background of thermal radiation.
The temperature of the background-radiation should be chosen to
correspond with the Hawking temperature
$T_\BH := \frac{1}{4\pi} \frac{1}{r_\BH}$
for the equilibrium of the system,
where $r_\BH$ is the Schwarzschild radius of the black hole.

The local proper temperature ansatz means that
the temperature-distribution of the radiation around the black hole
obeys
\begin{eqnarray}
 T(r) = \frac{T_\BH}{\sqrt{F(r)}}.
  \label{temperature.eq}
\end{eqnarray}
We expect that the radiation which consists from interacting particles
in the gravitational potential obeys the ansatz
\cite{Hotta:1997yj,Iizuka:2003ad}.
The analytic continuation of the angular-momentum distribution of the
Hawking radiation without the particle-interactions
also reproduces the ansatz \cite{Nagatani:2003ps}.
The energy-momentum tensor of the radiation becomes
\begin{eqnarray}
 &&
  \rho(r) \ =\  \frac{\pi^2}{30} g_* T^4(r), \qquad
  P(r) \ = \ \frac{1}{3} \rho(r),\nonumber\\
 &&
  T_\rad{}_\mu^{\ \nu} \ =\  \diag\Bigl[\, \rho, -P, -P, -P \, \Bigr],
\end{eqnarray}
where $g_*$ is the massless degree of freedom of the theory.
We should introduce the positive cosmological constant
$\Lambda = (8\pi/m_\pl^2) \rho_\vac$
to stabilize the background universe from the background energy density
$\rho_\BG := \frac{\pi^2}{30} g_* T_\BH^4$.
The background space-time becomes the Einstein static universe
of radiation
by choosing $\rho_\vac = \rho_\BG$.
By solving the Einstein equation
\begin{eqnarray}
 R_\mu^{\ \nu} -\frac{1}{2} R \delta_\mu^{\ \nu} - \Lambda\delta_\mu^{\ \nu}
 &=&
 \frac{8\pi}{m_\pl^2} T_{\rad}{}_{\mu}^{\ \nu},
 \label{Einstein.eq}
\end{eqnarray}
the radiation-ball solution is derived, where $m_\pl$ is the Planck mass.

We can numerically solve the Einstein equation \EQ{Einstein.eq}.
The Mathematica code for the numerical calculation
can be downloaded on \cite{MathCode:2003}
and the concrete forms of the numerical solution were shown
in \cite{Nagatani:2003rj,Nagatani:2003new}.
We find the solution whose exterior part ($r \gnear r_\BH + l_\pl$)
corresponds with the Schwarzschild solution.
We also find that the horizon $r = r_\BH$ of the black hole
vanishes and the radiation gravitationally trapped in the ball
of radius $r_\BH$ arises instead of the ordinary black hole.
This is the radiation-ball solution and it can be identified as the
ordinary black hole with the Hawking radiation.
%

When $r_\BH \gg l_\pl$,
we can approximately derive the analytic form of the solution
which is divided into three parts, namely,
the exterior of the ball $r \gnear r_\BH + l_\pl$,
the transitional region $r_\BH - l_\pl \lnear r \lnear r_\BH + l_\pl$
and the interior $r \lnear r_\BH - l_\pl$.

The metric elements of the exterior part ($r \gnear r_\BH + l_\pl$)
become
\begin{eqnarray}
 F_\external(r)
  &=& 1 -    \frac{r_\BH}{r},
  \label{Fexternal}\\
 G_\external(r)
  &=&
  \left[
   F_\external(r)
      + \frac{8\pi}{3 m_\pl^2} r^2 \rho_\BG
	\left\{F_\external^{-1}(r) - 3 F_\external(r)\right\}
  \right]^{-1}. \label{Gexternal}
\end{eqnarray}
The metric is identified as the Schwarzschild black hole
in the Einstein static universe of radiation.
The temperature distribution becomes
\begin{eqnarray}
  T_\external(r)
   &=&
   T_\BH  \times
   \left(1 - \frac{r_\BH}{r}\right)^{-1/2}.
\end{eqnarray}
Therefore the exterior of the ball is identified as the ordinary black
hole with Hawking radiation in 
the Einstein static universe of the radiation.

For the interior ($r \lnear r_\BH - l_\pl$) of the ball the elements
of the metric become
\begin{eqnarray}
 F_\internal(r)
  &=&
  \frac{g_*}{720 \sqrt{2\pi}}
  \frac{1}{m_\pl^2 r_\BH^2}
  \left( \frac{r_\BH}{r} \right)
  \left[1 - \left(\frac{r}{r_\BH}\right)^5\right]^{-1},\\
 G_\internal(r)
  &=&
   \frac{g_*}{72}
   \frac{1}{m_\pl^2 r_\BH^2}
   \left(\frac{r}{r_\BH}\right)
   \left[1 - \left(\frac{r}{r_\BH}\right)^5 \right]^{-3}
\end{eqnarray}
and the temperature distribution of the radiation becomes
\begin{eqnarray}
 T_\internal(r)
  &=&
  \frac{3 \sqrt{5} \cdot 2^{1/4}}{\pi^{3/4}} 
  \frac{m_\pl}{g_*^{1/2}}   
  \left(\frac{r}{r_\BH}\right)^{1/2}
  \left[ 1 - \left(\frac{r}{r_\BH}\right)^5 \right]^{1/2}.
\end{eqnarray}
The temperature $T_\internal(r)$ has the maximum value
$T_\maxi = (5 \cdot 3^{2/5})/(2^{7/20} \pi^{3/4}) (m_\pl/\sqrt{g_*})
\simeq 2.580 \times m_\pl/\sqrt{g_*} $
on the radius $r_{\rhopeak} = 6^{-1/5} r_\BH \simeq 0.6988 \times r_\BH$.
On the same radius, $F(r)$ has the minimum value
$
F_\mini =
g_* / (200 \cdot 2^{3/10} \cdot 9^{4/5} \sqrt{\pi} m_\pl^2 r_\BH^2)
\simeq 9.515 \times 10^{-4} g_* /(m_\pl^2 r_\BH^2)$.
Therefore the temperature of the radiation in the ball
is on the scale of the Planck energy.
The leak of the inside radiation is regarded as the Hawking radiation.

By the thermodynamical relation of the entropy
$s(T) = \frac{2\pi^2}{45} g_* T^3$,
the total entropy of the radiation in the ball is calculated as
\begin{eqnarray}
 S_\internal
  &\simeq&
  \int_{0}^{r_\BH} 4 \pi r^2 dr \sqrt{G_\internal(r)} \;
  s\left(T_\internal(r)\right)
  \ =\ 
  \frac{(8 \pi)^{3/4}}{\sqrt{5}} m_\pl^2 r_\BH^2
 \;\simeq\; 5.0199 \times \frac{r_\BH^2}{l_\pl^2}.
 \label{entropy-result.eq}
\end{eqnarray}
The entropy \EQ{entropy-result.eq}
is proportional to the surface-area of the ball
and
is a little greater than the Bekenstein entropy \cite{Bekenstein:1973ur}:
$ S_{\rm Bekenstein}
  \;=\; \frac{1}{4} \frac{\rm (Horizon\ Area)}{l_\pl^2}
  \;=\; \pi \times \frac{r_\BH^2}{l_\pl^2,}
$.
The ratio becomes $S_\internal/S_{\rm Bekenstein} \simeq 1.5978$.
Therefore the black hole entropy
is regarded as the total entropy of the radiation in the ball.

\section{Baryon Number Violation}\label{baryon.sec}

In the radiation-ball description of the black hole,
the baryon number is carried by the internal radiation of the ball
in the same way of the entropy.
Therefore we can define the baryon number density $b(r)$
and the total baryon number of the ball as
\begin{eqnarray}
 B &:=&
  \int_{0}^{r_\BH} 4 \pi r^2 dr \sqrt{G_\internal(r)} \; b(r)
 \label{B.eq}
\end{eqnarray}
which is regarded as {\it the baryon number of the black hole}.
This is one of the interesting features of
the radiation-ball description of the black hole.

The Higgs scalar vev $\left<\phi\right>$ in the radiation-ball vanishes
and the symmetry of the electroweak (EW) theory restores
because the temperature in the ball ($\sim$ Planck scale)
is much greater than the EW scale ($\sim100\GeV$)
and the thermal phase transition of the EW theory arises.
Therefore the sphaleron processes are working in the ball
and are changing the baryon number.
In the EW symmetric phase,
the sphaleron transition rate at temperature $T$ is given by
\begin{eqnarray}
 \Gamma_\sph = \kappa \alpha_\weak^4 T^4,
 \label{GammaSph}
\end{eqnarray}
where $\kappa\sim O(1)$ is a numerical constant and
$\alpha_\weak$ is the weak gauge coupling constant \cite{Arnold:1996dy,Huet:1996sh,Moore:1997sn}.
We have assumed no GUT and $\alpha_\weak$ is the coupling constant 
at the Planck energy.
We should note that the rate \EQ{GammaSph} is defined for the proper-time
rather than the coordinate time $t$.
We should take account of the red-shift effect in the ball
to consider the rate for the coordinate-time.
%
%
The time-evolution of the baryon number density
on the coordinate-time is given
by the Boltzmann-like equation:
\begin{eqnarray}
 \frac{d}{dt}b &=&
  \sqrt{F}
  \left[
    \frac{\Gamma_\sph}{T} \mu_B
  - \frac{39}{2}\frac{\Gamma_\sph}{T^3} b
  \right],
  \label{Boltzmann.eq}
\end{eqnarray}
where $\mu_B$ is the chemical potential for the baryon number
\cite{Garcia-Bellido:1999sv}.
The hyper-charge density of the radiation makes
the finite chemical potential
\begin{eqnarray}
 \mu_B &=& N \frac{q_Y}{T^2},
\end{eqnarray}
where $N \sim O(1)$ is a model-dependent constant and $q_Y$ is the
hyper-charge density of the radiation \cite{CKN}.
By substituting the temperature relation \EQ{temperature.eq},
\EQ{Boltzmann.eq} becomes
\begin{eqnarray}
 \frac{d}{dt}b &=&
  \kappa \alpha_W^4 T_\BH \left[ N q_Y - \frac{39}{2} b \right].
  \label{Boltzmann2.eq}
\end{eqnarray}

We introduce the total hyper-charge of the ball as
\begin{eqnarray}
 Q_Y &:=&
  \int_{0}^{r_\BH} 4 \pi r^2 dr \sqrt{G_\internal(r)} \; q_Y(r).
 \label{QY.eq}
\end{eqnarray}
It is natural to assume that the hyper-charge $Q_Y$ is much smaller
than the extremal charge $Q_Y^* := \alpha_Y/(m_\pl r_\BH)$ of the black
hole because
black holes loose their charges spontaneously
due to the electric repulsion and cannot have huge charges
\cite{Gibbons:1975kk,Hiscock:1990ex,Gabriel:2000mg}.
The time-evolution of the total baryon number becomes
\begin{eqnarray}
 \frac{d}{dt}B &=&
  R_{\rm BV}
  \left[ \frac{2 N}{39} Q_Y - B \right],
  \label{Boltzmann3.eq}
\end{eqnarray}
where we have defined the decay rate of baryon number of the radiation-ball:
\begin{eqnarray}
 R_{\rm BV}
  &:=&  \frac{39}{2} \kappa \alpha_W^4 T_\BH
  \ =\  \frac{39}{8\pi} \kappa \alpha_W^4 \frac{1}{r_\BH}
  \ \simeq\ \frac{10^{-6}}{r_\BH}.
  \label{RBV}
\end{eqnarray}
For the solar-massive black holes $r_\BH \simeq 1 {\rm km}$,
which are created by the gravitational collapse of stars
and are formed by huge baryonic matter,
the inverse of the rate $R_{\rm BV}^{-1}$ becomes order of seconds.
Such a black hole cannot obtain macroscopic hyper-charge.
Therefore we conclude that the baryon number of the black holes is
rapidly washed-out by the sphaleron processes in the ball.

When we assume that the GUT interactions also violate the baryon number,
the GUT process dominates over the sphaleron process
because the GUT process is perturbative and the rate of the GUT process
becomes $\Gamma_\GUT \approx \alpha_\GUT^2 T^4$. 
In the case the baryon-number decay-rate becomes
\begin{eqnarray}
 R_{\rm BV}
  &\simeq&  \frac{39}{8\pi} \alpha_\GUT^2 \frac{1}{r_\BH}
  \ \simeq\ \frac{10^{-3}}{r_\BH}.
  \label{RBV-GUT}
\end{eqnarray}
The rate becomes order of millisecond
for the solar-massive black holes.

\section{Conclusion and Discussion}\label{discussion.sec}

The radiation-ball description of black holes
allow us to consider time-evolution of the baryon-number
of the black holes concretely.
Because the entropy of the black hole is carried by the inside radiation
of the ball, the baryon number of the black hole is also carried by
the radiation and we can define the the total baryon number of the black
hole as \EQ{B.eq}.
The sphaleron process in the Standard Model violates the baryon number
at least. We can assume the baryon number violating process of the GUT.
The time-evolution of the baryon number of the black hole is described
by \EQ{Boltzmann3.eq}.
We find that the baryon number of the black hole is exponentially
decrease when the hyper-charge of the black hole is small.
The decay-rate of the baryon number without the GUT process is
evaluated as \EQ{RBV} and
the rate including the GUT process becomes \EQ{RBV-GUT}.
The rates depend on the Schwarzschild radius
and on the coupling constants of the relevant processes.
The time scale of the decay-rates is much shorter
than the lifetime of the black hole
$\tau_\BH = 1280\pi g_*^{-1} m_\pl^2 r_\BH^3$.
For the solar-massive black holes which are formed by the collapse of
the stars including huge baryonic matter,
the decay-rates become order of seconds without the GUT process and
order of milliseconds with the GUT process.
Therefore we conclude that
the baryon number of the matter
which has formed the black hole or has fallen into the black hole
rapidly decays.

The Hawking radiation on the radiation-ball description
is regarded as a leak-out of the inside radiation of the ball.
Therefore the radiation from the black holes includes
the net baryon number flux
when the black holes have nonzero baryon number \EQ{B.eq} and their
Hawking temperature is greater than mass of the proton $\sim 1\GeV$ as
the lightest baryon.
If we throw strong proton beams into such a small black hole,
we observe the flux of the net baryon number in the Hawking radiation.
When we stop the proton beams,
we find the baryon number flux is exponentially dumping
according to \EQ{Boltzmann3.eq}.

We also find that the hyper-charged black holes produce net baryon number.
If the hyper-charge $Q_Y$ of the black hole is constant,
the radiation-ball tries to keep its the baryon number to
$B = \frac{2 N}{39} Q_Y$ due to \EQ{Boltzmann3.eq}.
This is a kind of the homeostasis.
When the baryon number is taken away from the ball by the Hawking radiation,
the mechanism produces the net baryon number to cover the loss.
By combining the mechanism of the spontaneous charging-up of the black
hole \cite{Nagatani:2003pr},
we expect that the baryon number is spontaneously radiated
by the Hawking radiation.
This mechanism is similar to
the electroweak baryogenesis by black holes
\cite{Nagatani:1998gv,Nagatani:2001nz}.

\begin{flushleft}
 {\Large\bf ACKNOWLEDGMENTS}
\end{flushleft}

 I would like to thank
 Ofer~Aharony, Micha~Berkooz,
 Nadav~Drukker, Bartomeu~Fiol, Peter~Fischer, Hikaru~Kawai,
 Barak~Kol, Joan~Simon and Leonard~Susskind
 for useful discussions.
 I am grateful to Kei~Shigetomi
 for helpful advice and also for careful reading of the manuscript.
 The work has been supported by
 the Koshland Postdoctoral Fellowship of the Weizmann Institute of Science.


\end{document}